\documentstyle{mn}

\begin{document}
\title[Superluminal Cyg X-3]{Comments on the superluminal motion in Cygnus X-3}
\author[Ogley, Bell Burnell and Newell]{R. N. Ogley\thanks{E-mail: R.N.Ogley@open.ac.uk}$^1$, S. J. Bell Burnell$^1$ and S. J. Newell$^2$\\
$^1$Department of Physics, The Open University, Milton Keynes, UK.\\
$^2$NRAL, Jodrell Bank, Macclesfield, Cheshire, UK.}

\maketitle
\begin{abstract}
Following the recent discovery that Cyg X-3 exhibits superluminal
motion, the implications of superluminal expansion and contraction are
investigated.  We propose that the effect is due to either a
propagating photon pattern or to outwardly moving shells illuminated
by an intense beam of radiation.
\end{abstract}


\section{Introduction}

Since its discovery in 1966 (Giacconi et al. 1966) Cyg X-3 has
remained one of the most unusual and enigmatic objects in the sky.  A
neutron star--Wolf Rayet binary system, Cyg X-3 exhibits the following
unusual properties: major radio flares at intervals of aproximately 18
months (Waltman et al. 1995); a 4.8 hour period making it the shortest
period high mass X-ray binary; and following a major flare, radio jets
which expand at 0.35$c$ (Spencer et al.  1986).

Recently Newell, Garrett \& Spencer (1996, hereafter NGS; Newell 1996)
published VLBA maps showing superluminal motion at a significantly
higher magnitude than previous superluminal velocities in the galaxy.
The system also displayed superluminal contraction.  We consider
models for this behaviour.

\section{Superluminal motion}

Superluminal expansion in one-sided quasars has been well documented and
we use the same notation here.  Consider a blob of material emitted at
a velocity $v$ at an angle $\theta$ to the line of sight; the blob
appears to travel normal to the line of sight with a velocity $v_{\rm
app}$ given by

\begin{equation}
v_{\rm app} = \frac{v\sin{\theta}}{1-\beta\cos{\theta}}
\end{equation}
where $\beta = v/c$.  For a given
$\beta$ the angle at which the maximum superluminal effect occurs is
given by $\cos{\theta_{\rm max}} = \beta$ and at this angle the
apparent velocity has a maximum of $\beta_{\rm app}({\rm max}) =
\gamma\beta$ where we have substituted $\gamma =
(1-\beta^{2})^{-1/2}$.

\section{Cygnus X-3 results}

Observations by NGS of Cygnus X-3 showed that it was undergoing
apparent superluminal expansion and contraction on both the major and
minor axes of an ellipse.  The observations include frames showing the
object at intermediate size.  Taking $\beta_{\rm app}({\rm max}) =
\beta_{\rm app}$ and using the $\beta_{\rm app}$ values quoted by NGS,
we find $\beta$ and $\gamma$ as reported in Table 1.  Using
$\beta_{\rm app}({\rm max})$ for $\beta_{\rm app}$ will tend to
underestimate the actual velocities.  The $\beta_{\rm app}$ values
reported by NGS may also be underestimates of the superluminal
velocities (see section \ref{offsetcentre}).  We assume that the
distance adopted by NGS for Cyg X-3 is not seriously in error.

\begin{table}
\begin{tabular}{lllccc}\hline
Flare& Motion&Axis & $\beta_{\rm app}$ & $\beta$ & $\gamma$ \\ \hline
1&Expansion  &Major & 2.45 $\pm$ 0.55 & 0.920 & 2.56 \\
 &           &Minor & 0.84 $\pm$ 0.09 & 0.579 & 1.23 \\
 &Contraction&Major & 2.97 $\pm$ 0.33 & 0.947 & 3.12 \\
 &           &Minor & 2.53 $\pm$ 0.32 & 0.926 & 2.65 \\
2&Expansion  &Major & 4.75 $\pm$ 0.42 & 0.979 & 4.88 \\
 &           &Minor & 2.32 $\pm$ 0.32 & 0.914 & 2.46 \\
 &Contraction&Major & 6.76 $\pm$ 0.72 & 0.989 & 6.86 \\
 &           &Minor & 2.53 $\pm$ 0.54 & 0.926 & 2.65 \\ \hline
\end{tabular}
\caption{Apparent and actual velocities along the axes of the ellipse}
\end{table}

Questions that arise are:

\begin{itemize}
\item{What is moving?  Is it material moving out and radiating, or is
it a pattern of radiation illuminating fixed material?}
\item{Do any realistic models produce contraction?}
\item{Why are the speeds of expansion and contraction different, with
contraction being the faster?  Why does the second flare give greater
speeds?}
\item{Why is the shape elliptical and not circular?}
\end{itemize}

\section{Cygnus X-3 models}
We describe a variety of models which attempt to explain the
superluminal expansions and contractions shown in Table 1 along with
the observed elliptical shape.  A jet of radio-emitting plasma, an
expanding radiating shell, a beam of highly energetic particles or
radiation striking a stationary medium and exciting it, and a pattern
of radiation are considered.  We do not attempt to explain how such
conditions arise, nor do we consider radiation or cooling mechanisms.

\subsection{Bipolar jet model}

Superluminal expansion has been observed in the galactic bipolar jet
sources GRS 1915+105 (Mirabel \& Rodr\'{\i}guez 1994) and GRO 1655-40
(Tingay et al. 1995).  Although the expansion velocities observed here
are much larger, could such a model explain the superluminal expansion
of Cyg X-3?  To explain superluminal {\it contraction} however by a
similar process the bulk motion of the material would have to be
reversed, which is unrealistic.  Nor does this model account readily
for the elliptical shape of the emission.  It is worth noting that
while this model does not account for the motion in Cygnus X-3, it is
consistent with the relativistic motions in GRS 1915+105 and GRO
J1655-40 and cannot be discounted for these sources.

Before moving to the next model, note that an apparently superluminal
blue-shifted jet is accompanied by an apparently subluminal
red-shifted jet.  In obtaining their values for $\beta_{\rm app}$ NGS
assumed that the apparent expansion (or contraction) was from (or to)
a central point, and of equal magnitude in opposite directions.  Thus
the superluminal speeds have probably been underestimated.  This is
developed further in the next section.

\subsection{Offset centre}
\label{offsetcentre}
A consequence of the hybrid mapping technique used by NGS is that
absolute positional information is lost and so it is not possible accurately to locate the ellipses relative to each other or to the core of
the system. NGS assume that the bright core of each map represents the
same feature.

If the superluminal expansion speeds along the major axis are caused
by a bulk motion with a component towards the observer, then the speed
in the opposite direction must be subluminal, and the point from which
the expansion takes place must be offset from the observed centre of
the major axis.  The same argument applies to motion along the minor
axis.  The source of the expanding material is located at or near the
rim of the ellipse, lying on neither its major nor minor axis, and the
superluminal expansion speeds are thus approximately doubled (Table 2)

An outburst might produce an elliptical lobe with the central source
lying at one end of the major axis of the ellipse, but seems unlikely
to do so for the central source offset from both the major and minor
axes.  In addition this model does not readily explain the
superluminal contraction.  We set this model aside.

\begin{table}
\begin{tabular}{lcccc}\hline
Axis & $\beta_{\rm app}$ & $\beta^{\prime}_{\rm app}$ & $\beta^{\prime}$ &
$\gamma^{\prime}$ \\ \hline 
Major & 4.8 & 9.6 & 0.995 & 9.64 \\ 
Minor & 2.3 & 4.6 & 0.977 & 4.68 \\ \hline
\end{tabular}
\caption{Projected velocities and true velocities based on a shifted expansion centre based on flare 2.}
\end{table}

\subsection{Christmas tree model or propagating photon pattern}

The velocities found in the previous models are at the limit of
physical reality.  If, however, the superluminal effect is caused by a
pattern of photons propagating, then relativistic bulk motion, and the
reversal of relativistic bulk motion may not be necessary.  Equation
1, with $v=c$ applies.

If an intense burst of radiation was emitted from the core of Cyg X-3
the observer might, in addition to seeing some of these photons
directly, see either photons reflected/scattered off surrounding
material or secondary photons generated in this material following
excitation by the burst of radiation.  The size of the emitting patch
would be governed by the extent of the distribution of material around
the core of Cyg X-3, by the distance from the core that the burst
of radiation had travelled and possibly by the excitation time for
the case of secondary photons.

It is possible to envisage an expanding patch of emission, growing as
the burst of radiation travels further out.  The maximum size observed
is determined by the duration of the burst or by the extent of the
surrounding material or (less likely) by the optical depth of the
surrounding material to the centrally emitted radiation.  Note however
that equation 1 is cylindrically symmetric about the line of sight, so
that an isotropic burst of radiation into an isotropic distribution of
surrounding material would give a circular patch of emission.  The
observed elliptical shape could be produced if either the burst of
radiation or the surrounding material were confined to a disc inclined
to the line of sight.  Equation 1 requires that the total apparent
expansion speed along any axis is $\geq$ 2$c$, and the observations
satisfy this.  Along the axis of the ellipse that lies in the plane of
the sky the total apparent expansion speed takes the minimum value of
2$c$.  With higher signal-to-noise observations it should be possible
to identify this axis and determine the orientation of the disc in
space.

The superluminal contraction, in this model, is most likely explained
by a steady reduction in the effective extent of the photon pattern.
Either the central intensity drops continuously and insufficient
radiation reaches the outer areas to make visible the material there,
or (less likely) a steady change of the central wavelength makes the
optical depth gradually greater.  It is unlikely that the extent of
the surrounding material shrinks superluminally.  Another possibility
is that we are seeing the cooling of an excited region after the
central radiation has turned off.  However if the radiation ceases
totally, the central parts of the patch cool first.  This is not what
is observed.

\subsection{Searchlight beam model}
\label{radiation}

If a conical jet of high energy photons or particles was ejected from
the Cyg X-3 core, and there was some absorption by material in a
surrounding stationary spherical shell, then that part of the shell
intersected would become excited and radiate.  If the conical beam is
inclined to the line of sight then the observed emission patch has an
elliptical shape.  For flare 2 the required eccentricity is produced
if the cone axis is at an angle of 61$\degr$ to the line of sight.  If
the inclination varies by $\sim 10\degr$ then the observed changes in
eccentricity can be accommodated.  The observed emission is centrally
peaked suggesting that the beam also is more intense along its central
axis.

The apparent superluminal expansions and contractions could be
produced by relatively small expansions and contractions of the
opening angle of the cone, provided that the heating and cooling time
constants are less than the timescales for the change of cone angle.
The confinement of the beam might be by a magnetic throat, with the
opening angle in part governed by the flux in the beam,
producing the observed correlation beween size and intensity of the
emission.

Alternatively, at least a part of the expansion and contraction could
be produced by the weaker emission from the edge of the ellipse rising
above and falling below a detection threshold.

\subsection{Doppler boosted spot on an expanding shell}
\label{spotprojection}

\begin{figure}
\begin{picture}(10,140)
\put(-85,-510){\includegraphics{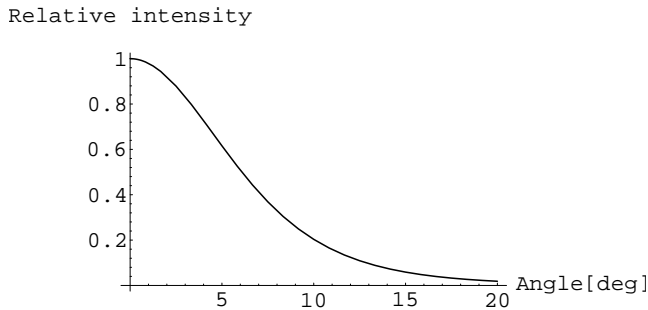}}
\end{picture}
\caption{Plot of Doppler boosted intensity against offset angle.  For a dynamical range of 32, the maximum angle at which flux is detected is 17.6 degrees.  The plot is for the major axis of the second flare; $\gamma = 4.88$.}
\label{Dopplerboosting}
\end{figure}

Consider a spherical shell of material expanding relativistically at
fixed velocity $V$.  The emission from material travelling at small
angles, $\phi$, to the line of sight will have its intensity Doppler
boosted according to

\begin{equation}
I(\phi) = \frac{I^{\prime}(\phi)}{\gamma^{3}\left(1 - \beta\cos{\phi}\right)^{3}}
\end{equation}
where $I^{\prime}(\phi)$ is the intensity in the rest frame of the radiating material (Rybicki \& Lightman, 1979).
Assuming $I^{\prime}(\phi)$ to be constant, a plot of observed
intensity with angle is shown in Figure \ref{Dopplerboosting}.  The
shell appears to have a bright spot centred on our line of sight,
which expands as the shell expands.

\begin{figure*}
\begin{picture}(60,120)
\put(-200,0){\includegraphics{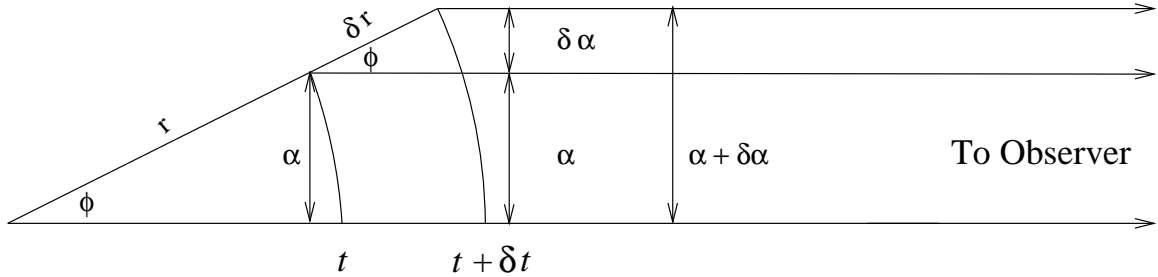}}
\end{picture}
\caption{Geometries for a spherically-symmetric shell expanding at a fixed velocity, $V$.  The observed angular radius of a spot on the shell as it expands is shown by the values $\alpha$, $\alpha + \delta\alpha$.}
\label{expandingshells}
\end{figure*}

Let $r$ be the radius of the shell at time $t$, and $\alpha$ the angular diameter of the spot (Figure \ref{expandingshells}).
The shell is expanding at a velocity $V = \delta r / \delta t$ and the
observed expansion of the spot can be written as $v = \delta\alpha /
\delta t$. Therefore $\sin{\phi} = \alpha / r = \delta\alpha / \delta r$, hence
$V = \left(r / \alpha\right)v$ giving $v < V$.  To find $(r/\alpha)$,
we use the dynamical range of the {\it VLBA} images to give the limit
at which lack of Doppler boosting is unable to bring the intensity up
to detectable levels.

The dynamical range for the VLBA images in Cyg X-3 is 32:1 which would
imply from Figure \ref{Dopplerboosting} that any flux outside the
angle of 17.6$\degr$ is undetectable.  The profile used for this assumes
$\gamma = 4.88$.  If $\phi = 17.6\degr$ then $(r/\alpha) = 3.30$ which
implies shell speeds for the second flare as shown in Table 3.

\begin{table}
\begin{tabular}{lcccc}\hline
Axis & $\beta_{\rm app}({\rm spot})$ & $\beta_{\rm app}({\rm shell})$ & $\beta(\rm shell) $ & $\gamma({\rm shell})$ \\ \hline
Major & 4.8 & 15.8 & 0.998 & 15.8 \\
Minor & 2.3 & 7.58 & 0.991 & 7.63 \\ \hline
\end{tabular}
\caption{Adjusted velocities for a shell of material with a Doppler boosted spot}
\label{shellspeeds}
\end{table}

In this model a spherical shell will produce a circular spot with
circular expansion velocities.  To simulate the elliptical shape
observed we postulate a shell distorted by expansion into an
anisotropic dense medium.

A dense medium, sufficient to decelerate the shell, is required to
explain the contraction.  In this model the apparent contraction is
due to a rapid reduction in intensity, lowering the emission from most
of the spot area below the detection threshold.  The rapid reduction
in intensity is due to the collapse of the Doppler boosting when the
expansion velocity drops.

A reduction in Doppler boosting of the intensity of 1/32 would require
a velocity change of $\delta\beta = - 0.819$.  If this occurs in
$\sim$ 60 minutes (the typical time between contractions in the NGS
observations) then the retardation is $\sim$ 70 km s$^{-2}$ and the
shell pushes back the retarding medium by $\sim$ 900$R_{\odot}$.
Considerable energy would be transfered to the medium, presumably with
detectable consequences.

\subsection{Illuminated shells}
\label{illuminatedshell}

This hybrid model, which combines the useful features of the moving
shell and the searchlight beam models ($\S\S$ 4.4 and 4.5) best
addresses the difficult questions concerning the contraction and the
elliptical shape of the emitting area.

The central source produces a series of expanding spherical shells,
and a beam of energetic particles or photons.  The beam, which is
inclined to the line of sight, illuminates a patch on an expanding
shell which is seen as an elliptical area of emission.  As the shell
expands the area expands superluminally, as set out in $\S$ 4.5, with
the minor axis expansion velocities apparently smaller than the major
axis ones.  Doppler boosting of the intensity is not significant here
because of the large angle to the line of sight.

However, as the shell expands the illuminating beam intensity per unit
area decreases and so the emitted radiation falls.  If the intensity
of the emission across the area has a flat distribution and is close
to the detection threshold of the observer's equipment then as the
source fades its detectable area will rapidly shrink simulating
superluminal contraction.

Meanwhile another shell has been produced and is expanding in the wake
of the first.  As emission from the first shell fades, the expanding
spot on this second shell becomes visible.

\begin{figure*}
\begin{picture}(60,250)
\put(-150,0){\includegraphics{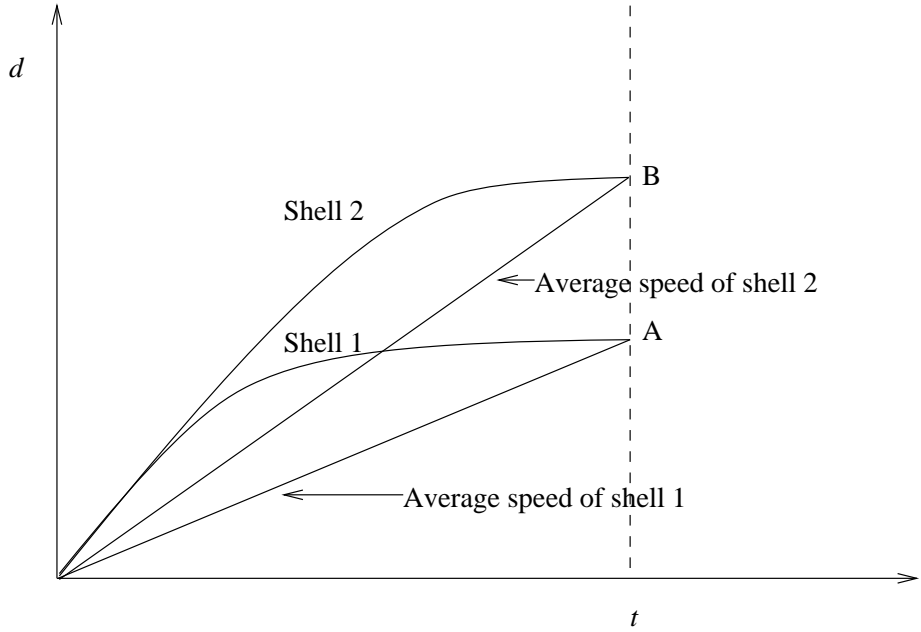}}
\end{picture}
\caption{Shell expansions based on a constant initial speed for two shells.  Shell 1 travels to a distance A before it is halted by some braking medium.  Flare 1 moves the braking medium back a bit so shell 2 travels out further before being decelerated and thus has a higher average speed.  The distance between A and B is approximately 900 $R_{\odot}$.}
\label{shellvel}
\end{figure*}

The difference in apparent speeds for the two flares can be explained
if we imagine the flares to be running into some ambient medium and
imparting momentum.  A graph of the distance travelled by the shell
against time is shown in Figure \ref{shellvel}.  If both shells are
expanding at the same initial rate, we would observe flare 1 to
travel unhindered, then run into an object that slows it down.  After
a time $t$ we would have observed it to have travelled a distance $A$
at an average velocity $v_{1} = A/t$.  During this deceleration, if
the shell pushes back the ambient medium it will have allowed flare 2
to expand a greater distance before being decelerated.  For the second
flare the initial speed is the same, but the shell travels further, to
$B$ before it is slowed down.  On the same time-scale, $t$, the
average speed will have increased to $v_{2} = B/t$.  The maximum shell
expansion speed occurs when a previous shell has pushed the
braking medium out far enough so we do not see deceleration within
the time scale $t$.  This is the true expansion velocity of the shell.

In this model a shell apparently expanding at 16$c$ produces a
superluminal expansion of an elliptical area which then fades and
appears to contract superluminally.  The model accounts for the
elliptical shape and the lower minor axis velocities.  Illumination of
a subsequent shell produces the next expansion and contraction phase.
This shell is expected to travel further before retardation and have a
higher average speed.  

\section{Conclusions}

We have considered a number of models which attempt to explain the
superluminal expansions and contractions observed in Cygnus X-3, and
the observed elliptical shape.  The latter two features are the most
difficult to model.  The two most successful models are a) the
propagating photon pattern model and b) the model in which expanding
shells of material are illuminated by an off-axis jet or beam of
radiation.

\section*{Acknowledgements}

We would like to thank Pete Taylor and Tim Ash for useful comments
during the development of the models presented here.  RNO and SJN
acknowledge the support of PPARC studentships.

\end{document}